\documentclass[twocolumn,english]{revtex4-1}
\usepackage[T1]{fontenc}
\usepackage[latin9]{inputenc}
\usepackage{float}
\usepackage{amsmath}
\usepackage{amsthm}
\usepackage{amssymb}
\usepackage{graphicx}
\usepackage{wasysym}

\makeatletter

\usepackage{braket}
\usepackage{amsmath}
\DeclareMathOperator{\Tr}{Tr}

\makeatother

\usepackage{babel}
\begin{document}

\title{Multipartite steering inequalities based on entropic uncertainty
relations}

\author{Alberto Riccardi, Chiara Macchiavello and Lorenzo Maccone}

\affiliation{Dip. Fisica and INFN Sez. Pavia, University of Pavia, via Bassi 6,
I-27100 Pavia, Italy}
\begin{abstract}
We investigate quantum steering for multipartite systems by using
entropic uncertainty relations. We introduce entropic steering inequalities
whose violation certifies the presence of different classes of multipartite
steering. These inequalities witness both steerable states and genuine
multipartite steerable states. Furthermore, we study their detection
power for several classes of states of a three-qubit system. 
\end{abstract}
\maketitle
Quantum steering is a type of quantum correlation, owned by some entangled
states of composite systems. It enables one subsystem to influence
the state of the others, with which it shares the entangled state,
by applying local measurements. The concept of quantum steering, for
bipartite systems, was introduced in the early days of quantum mechanics
by Schrodinger \cite{Sch}, who recognized that this class of states
allow one part ``to steer'' the state of the other into an eigenstate
of an arbitrary observable, and hence they express the ``spooky action
at distance'' discussed in \cite{EPR}. Nowadays we are aware that
three types of quantum entanglement exist: Bell nonlocality, steerability
and nonseparability. Bell nonlocality correlations are the strongest
ones and are owned by global states that violate some Bell inequalities
\cite{Bell}, which are related to the non existence of local hidden
variable (LHV) models. Then we have quantum steering, which was formalized
in 2007 by Wiseman \emph{et.al} \cite{Wiseman} as the incompatibility
of quantum mechanics predictions with a local hidden state (LHS) model,
where the parties have pre-determined states. Formally, given a bipartite
system owned by Alice and Bob that share a state $\rho^{AB}$ , we
say that the correlations demonstrate quantum steering if the joint
measurement probabilities cannot be expressed as:
\begin{equation}
p\left(x_{a},x_{b}\right)=\varint d\lambda q\left(\lambda\right)p\left(x_{a}|\lambda\right)p_{\lambda}\left(x_{b}\right),\label{Bipartite LHS model}
\end{equation}
where $x_{a}$ and $x_{b}$ are respectively the outcomes of the measurements
of Alice's observable $X_{A}$ and Bob's observable $X_{B}$. In the
above equation $p_{\lambda}\left(x_{b}\right)$ represents the probability
of $x_{b}$ obtained from a quantum pre-determined state $\rho_{\lambda}^{B}$
that depends only on $\lambda,$which occurs with probability $q\left(\lambda\right)$
and not on $x_{a}$. Instead, the conditional probability of $x_{b}$
on an arbitrary state, which may depend on $x_{a}$, will be indicated
as $p\left(x_{b}|\lambda\right).$ Conversely if for any choice of
measurements equation (\ref{Bipartite LHS model}) holds, then the
state is called nonsteerable, in the sense that it admits a LHS model.
\\
At the bottom of the hierarchy there is entanglement \cite{Hor,Wer1},
which can be defined as the existence of states of composite systems
that cannot be given as a convex combination of states of the individual
subsystems, namely separable states. Interestingly, these three notions,
which can be only found in nonseparable states, coincide for pure
states.\\
All of these types of correlations have been generalized to multipartite
systems. However for steerability there exist different approaches
\cite{HeReid,Multisteering1,Indian} that go beyond the bipartite
scenario. Here we consider the one discussed in \cite{Multisteering1},
which also allows one to discuss the notion of post-quantum steering
\cite{Postquantumsteering}, which does not exist for bipartite systems.\\
In this paper we introduce a number of entropic inequalities whose
violation certifies multipartite steering. Steerability is an asymmetric
concept, i.e. one part steers the others. In multipartite systems
there exist several different steering scenarios, depending on how
many subsystems steer the others. For example in a tripartite system
we can have one subsystem that tries to steer the other two, a scenario
that we indicate as\emph{ one-to-two steering}, or two subsystem that
might steer the other one, which we refer as\emph{ two-to-one steering}.
As in the case of entanglement we have different levels of multipartite
steerability \cite{Multisteering1}. \\
In the \emph{one-to-two steering} scenario we say that the correlations
demonstrate multipartite steering \cite{Multisteering1} if the joint
measurement probabilities cannot be expressed as:\\
\begin{equation}
p\left(x_{a},x_{b},x_{c}\right)=\varint d\lambda q\left(\lambda\right)p\left(x_{a}|\lambda\right)p_{\lambda}\left(x_{b}\right)p_{\lambda}\left(x_{c}\right),\label{A-BC LHS non steering}
\end{equation}
where $x_{a},x_{b}$, and $x_{c}$ are the outcomes the observables
$X_{A},X_{B}$ and $X_{C}$ of Alice, Bob and Charlie respectively.
In (\ref{A-BC LHS non steering}) Bob and Charlie's quantum state
is pre-determined, for a given $\lambda$ their state is $\rho_{\lambda}^{B}\otimes\rho_{\lambda}^{C}$.
A state is said instead to demonstrate genuine multipartite steering
\cite{Multisteering1} if the joint measurement probabilities cannot
be written as:
\begin{align}
p\left(x_{a},x_{b},x_{c}\right)= & \varint_{A}d\nu q_{A}\left(\nu\right)p\left(x_{a}|\nu\right)p_{\nu}\left(x_{b},x_{c}\right)\label{A-BC NON-GMS}\\
 & +\varint_{B}d\gamma q_{B}\left(\gamma\right)p\left(x_{a}|\gamma\right)p_{\gamma}\left(x_{b}\right)p\left(x_{c}|x_{a},\gamma\right)\nonumber \\
 & +\varint_{C}d\omega q_{C}\left(\omega\right)p\left(x_{a}|\omega\right)p\left(x_{b}|x_{a},\omega\right)p_{\omega}\left(x_{c}\right),\nonumber 
\end{align}
where $q_{A}\left(\nu\right),q_{B}\left(\gamma\right)$ and $q_{C}\left(\omega\right)$
satisfy: $\int_{A}d\nu q_{A}(\nu)+\int_{B}d\gamma q_{B}(\gamma)+\int_{C}d\omega q_{C}(\omega)=1.$
In (\ref{A-BC NON-GMS}) there are three terms: in the first there
is no steering between Alice, Bob and Charlie, in the second Alice
can steer Charlie but not Bob, namely only Bob's state is pre-determined
for a given $\gamma$, conversely in the third Alice can steer Bob
but not Charlie, which means that only Charlie's state is pre-determined
for a given $\omega.$\\
In the \emph{two-to-one steering} scenario we say that the correlations
demonstrate multipartite steering \cite{Multisteering1} if the joint
measurement probabilities cannot satisfy: 
\begin{equation}
p\left(x_{a},x_{b},x_{c}\right)=\varint d\lambda\mu\left(\lambda\right)p\left(x_{a}|\lambda\right)p\left(x_{b}|\lambda\right)p_{\lambda}\left(x_{c}\right).\label{AB-C NON-Steering}
\end{equation}
Conversely if the above holds the state is nonsteerable from Alice
and Bob to Charlie, indeed Charlie's state is pre-determined by the
value of $\lambda.$ In this scenario a state is said to be GMS \cite{Multisteering1}
if the joint measurement probabilities cannot be expressed as: 
\begin{align}
p\left(x_{a},x_{b},x_{c}\right)= & \varint_{A}d\nu q_{A}\left(\nu\right)p\left(x_{a}|\nu\right)p\left(x_{b}|\nu\right)p\left(x_{c}|x_{b},\nu\right)\nonumber \\
 & +\varint_{B}d\gamma q_{B}\left(\gamma\right)p\left(x_{a}|\gamma\right)p\left(x_{b}|\gamma\right)p\left(x_{c}|x_{a},\gamma\right)\nonumber \\
 & +\varint_{C}d\omega q_{C}\left(\omega\right)p\left(x_{a},x_{b}|\omega\right)p_{\omega}\left(x_{c}\right),\label{AB-C NON-GMS}
\end{align}
where $\int_{A}d\nu q_{A}(\nu)+\int_{B}d\gamma q_{B}(\gamma)+\int_{C}d\omega q_{C}(\omega)=1.$
. In (\ref{AB-C NON-GMS}) the first term shows that only Bob can
steer Charlie, in the second only Alice and in the third Alice and
Bob cannot jointly steer Charlie. However Alice and Bob can share
entanglement.

As any type of quantum correlations, one of the problems connected
with quantum steering is its detection. Several methods have been
introduced in the last years for detecting steering in bipartite systems,
for example \cite{Kogias,key-14,key-15,Walborn,EURSteering}. Here
we are interested in entropic steering criteria such as the one defined
in \cite{Walborn,EURSteering,JS2,Prak}. In \cite{Walborn} it was
derived that a nonsteerable state satisfies: 
\begin{equation}
H\left(X_{B}|X_{A}\right)\geq\int d\lambda q\left(\lambda\right)H_{\lambda}\left(X_{B}\right),\label{Walborn ineq}
\end{equation}
where $H\left(X_{B}|X_{A}\right)$ is the conditional entropy of $X_{B}$
given $X_{A}$ and $H_{\lambda}\left(X_{B}\right)$ is the conditional
entropy of $X_{B}$ computed on $\rho_{\lambda}^{B}$, that does not
depend on Alice's measurements. Thus any violation of (\ref{Walborn ineq})
demonstrates steering from Alice to Bob. In \cite{EURSteering} the
inequality (\ref{Walborn ineq}) was generalized to state-independent
entropic uncertainty relations (EUR). As an example, given any two
of Alice's observables $X_{A}$ and $Z_{A}$ and two of Bob's observables
$X_{B}$ and $Z_{B}$, for any nonsteerable state the following holds:
\begin{equation}
H\left(X_{B}|X_{A}\right)+H\left(Z_{B}|Z_{A}\right)\geq-\log_{2}\alpha_{B},\label{MUF steering}
\end{equation}
where $\alpha_{B}=\max_{j,k}\left|\braket{x_{j}^{B}|z_{k}^{B}}\right|^{2},$
with $\left\{ \ket{x_{j}^{B}}\right\} _{j}$ and $\left\{ \ket{z_{k}^{B}}\right\} _{k}$
the eigenstates of $X_{B}$ and $Z_{B}$ respectively. Eq. (\ref{MUF steering})
is a generalization of Maaseen and Uffink's EUR \cite{MUF} to nonsteerable
states, which can be violated only by steerable states from Alice
to Bob. Starting from (\ref{Walborn ineq}) other inequalities of
the form (\ref{MUF steering}) can be derived simply by considering
different EUR from the ones of \cite{MUF}, for example the ones derived
in \cite{EURreview,Ivanovi,RiccardiMM,RPZ,Sanchez,Karol1,key-16,KK,Li,Deut,ColesPiani}.
\\
In this paper we derive the following results: \\
$\left(i\right)$ we first show that (\ref{Walborn ineq}) and (\ref{MUF steering})
can be generalized to tripartite steering, obtaining different sufficient
conditions for both steerable and GMS states. In the case of\emph{
one-to-two steering }scenario where Alice, whose measurements are
uncharacterized, might steer Bob and Charlie's state, we show that
any nonsteerable state satisfies the following set of entropic uncertainty
relations: 

\begin{align}
\sum_{O=X,Z}H\left(O_{m}|O_{A}\right) & \geq-\log_{2}\alpha_{m};\label{A-BC NS EUR1}
\end{align}
where $\alpha_{m}=\max_{j,k}\left|\braket{x_{j}^{m}|z_{k}^{m}}\right|^{2}$
with $m=B,C,BC$ labeling the subsystem considered and $\left\{ \ket{x_{j}^{m}}\right\} _{j}$
and $\left\{ \ket{z_{k}^{m}}\right\} _{k}$ being the eigenstates
of $X_{m}$ and $Z_{m}$ respectively. Here $O_{BC}$ is given by
$O_{B}\otimes O_{C}$ for any observables. \\
$\left(ii\right)$ We also show that
\begin{equation}
\sum_{O=X,Z}H\left(O_{m}|O_{A}O_{\bar{m}}\right)\geq-\log_{2}\alpha_{m},\label{eq 9}
\end{equation}
holds for all nonsteerable states, where $m=B,C$ and $\bar{m}$ indicates
the opposite of $m$, i.e. $\bar{m}=B$ if $m=C$ and $\bar{m}=C$
when $m=B$. \\
$\left(iii\right)$The last inequality for nonsteerable states involves
the following quantity: 
\begin{equation}
A\left(O_{A},O_{B},O_{C}\right)=H\left(O_{BC}|O_{A}\right)+\sum_{m=B,C}H\left(O_{m}|O_{A},O_{\bar{m}}\right).
\end{equation}
We prove that for a nonsteerable state the following holds:
\begin{equation}
\sum_{O=X,Z}A\left(O_{A},O_{B},O_{C}\right)\geq-4\log_{2}\alpha_{min},\label{33-1}
\end{equation}
where $\alpha_{min}=\min\left\{ \alpha_{B},\alpha_{C}\right\} .$
\\
$\left(iv\right)$ For any non-GMS states we prove that the following
inequality is satisfied: 
\begin{equation}
\sum_{O=X,Z}A\left(O_{A},O_{B},O_{C}\right)\geq-2\log_{2}\alpha_{min}.\label{EUR Fin}
\end{equation}
\\
$\left(v\right)$ Finally, we give the following state-dependent entropic
uncertainty relation valid for all non-GMS states:
\begin{gather}
\sum_{O=X,Z}H\left(O_{BC}|O_{A}\right)\geq-\log_{2}\alpha_{CB}\nonumber \\
+\int_{B}d\gamma q_{B}\left(\gamma\right)S_{\gamma}\left(C|A\right)+\int_{C}d\omega q_{C}\left(\omega\right)S_{\omega}\left(B|A\right),\label{memory eurP-1}
\end{gather}
where $S_{\gamma}\left(C|A\right)$ represents the Von Neumann conditional
entropy between the bipartition $C|A$ when the variable $\gamma$
occurs, while $S_{\omega}\left(B|A\right)$ is the Von Neumann conditional
between Bob and Alice when $\omega$ occurs. Note that these quantities
can be negative \cite{Cerf} for entangled states, moreover their
lowest values $-\log_{2}d_{C}$ and $-\log_{2}d_{B}$ are reached
by maximally entangled states. We note that the inequality (\ref{memory eurP-1})
is not useful in the context of multipartite steering detection, if
one wants to understand the steering property of an unknown state,
since it requires the knowledge of the LHS model (\ref{A-BC NON-GMS}).
Conversely, the multipartite steering criteria (\ref{A-BC NS EUR1}-\ref{EUR Fin})
can be exploited in the task of discovering the steering property
of unknown quantum states. In order to compare the power in detecting
steerability of the criteria in Section V we consider the steerability
robustness of the standard $GHZ$ and $W$ states under white noise
and we show that criterion (\ref{eq 9}) detects more multipartite
steerable states than the others.\\
The results (\ref{A-BC NS EUR1}-\ref{EUR Fin}) are also extended,
with the same techniques, to the \emph{two-to-one} steering scenario.
\\
The paper is organized as follows: in section I we review bipartite
quantum steering by following the approach of \cite{Review}. Here
we also report the derivations of (\ref{Walborn ineq}) and (\ref{MUF steering}).
In section II we review the definition of multipartite steering, which
it was introduced in \cite{Multisteering1}. In section III we focus
on the \emph{one-to-two steering} scenario and we derive the steering
inequalities (\ref{A-BC NS EUR1},\ref{EUR Fin}). In section IV the
results for the \emph{two-to-one steering} are discussed. Finally
in section V some steering states are considered in order to study
the detection power of these inequalities. 

\section{Bipartite quantum steering}

\subsection{Definition and LHS model}

Bipartite quantum steering \cite{Wiseman} can be seen as the ability
to nonlocally influence the set of possible quantum states of a given
system through the measurements of another system sufficiently entangled
with the first one. In the steering scenario Alice and Bob share a
quantum state $\rho^{AB}$ and Alice performs a measurement $X_{A}$
whose outcome $x_{a}$ occurs with probability $p\left(x_{a}\right)$.
As a consequence of Alice's measurements, Bob's state is transformed
into the state $\rho_{x_{a}}^{B}$ with probability $p\left(x_{a}\right)$.
Here we do not require any characterization of Alice's measurements,
namely we only say that she performs an arbitrary measurement, and
we suppose that Bob has full access to the conditional state $\rho_{x_{a}}^{B}$
and on his measurements. Namely, the information available to Bob
is the collection of the post-measured states and their respective
probabilities $p\left(x_{a}\right)$, which can be described with
the following ensemble of unnormalized states: 
\begin{equation}
\left\{ \sigma_{x_{a}}^{B}=p\left(x_{a}\right)\rho_{x_{a}}^{B}\right\} .\label{Ensemble}
\end{equation}
Each member of (\ref{Ensemble}) is given by: 
\begin{equation}
\sigma_{x_{a}}^{B}=\Tr_{A}\left[\left(\Pi_{x_{a}}^{A}\otimes\mathbb{I^{B}}\right)\rho^{AB}\right],
\end{equation}
where $\sum_{x_{a}}\Pi_{x_{a}}^{A}=\mathbb{I}^{A}$ and $\Pi_{x_{a}}^{A}\geq0$
are Alice's POVM elements. The ensemble (\ref{Ensemble}) represents
the set of possible quantum states that can be nonlocally influenced
when steering correlations are owned by $\rho^{AB}$. Therefore the
LHS model formally represents the minimal requirement on (\ref{Ensemble})
in order to avoid this nonlocal influence, then steering is defined
as the possibility of remotely generating ensembles that could not
be produced by a LHS model. This model can be thought in the following
way: a source sends, according to a probability distribution $q(\lambda)$,
a classical message $\lambda$ to Alice, her probability of obtaining
$x_{A}$ depends now on $\lambda$: $p\left(x_{A}|\lambda\right)$.
To each $\lambda$ there corresponds a pre-determined state of Bob
$\rho_{\lambda}^{B}$, which is sent to Bob with the same probability
$q(\lambda).$ Bob's ensemble (\ref{Ensemble}), that now does not
depend on Alice's measurements, is given by: 
\begin{equation}
\sigma_{x_{A}}^{B}=\int d\lambda q(\lambda)p\left(x_{a}|\lambda\right)\rho_{\lambda}^{B}.\label{LHS Model state}
\end{equation}
The definition of steering is as follows: an ensemble (\ref{Ensemble})
is said to demonstrate bipartite steering if it does not admit a decomposition
of the form (\ref{LHS Model state}). Moreover a quantum state $\rho^{AB}$
is said to be steerable from Alice to Bob if there exists a measurement
in Alice's part that produces an ensemble that demonstrates steering.
This is an asymmetric concept that also implies entanglement. Indeed
suppose that $\rho^{AB}$ is separable, namely we have $\rho_{S}^{AB}=\int d\lambda q(\lambda)\rho_{\lambda}^{A}\otimes\rho_{\lambda}^{B}$.
After Alice has performed a measurement, Bob's ensemble becomes: 
\begin{align}
\sigma_{x_{a}}^{B}= & \Tr_{AB}\left[\left(\Pi_{x_{a}}^{A}\otimes\mathbb{I}^{B}\right)\rho_{S}^{AB}\right],\\
 & =\int d\lambda q(\lambda)\Tr_{A}\left[\Pi_{x_{a}}\rho_{\lambda}^{A}\right]\rho_{\lambda}^{B};\nonumber 
\end{align}
which is of the form (\ref{LHS Model state}). Since it implies nonseparability,
steering detection can be seen as an entanglement detection task where
one part, the one that steers, performs arbitrary measurements and
its system remains completely uncharacterized, namely we do not assume
anything on it, not even its dimension. The existence of a LHS model
can be written also in terms of joint probabilities of measurements,
namely by the condition (\ref{Bipartite LHS model}). Indeed we have:
\begin{align}
p\left(x_{a},x_{b}\right)=p\left(x_{b}|x_{a}\right)p\left(x_{a}\right)= & \Tr_{B}\left[\Pi_{x_{b}}^{B}\sigma_{x_{a}}^{B}\right]\\
= & \varint d\lambda q\left(\lambda\right)p\left(x_{a}|\lambda\right)p_{\lambda}\left(x_{b}\right).\nonumber 
\end{align}

\subsection{Entropic uncertainty steering inequalities }

Here we review the techniques used in \cite{Walborn} and \cite{EURSteering}
to derive (\ref{Walborn ineq}) and (\ref{MUF steering}). Suppose
that a state $\rho^{AB}$ admits a LHS model, then (\ref{Bipartite LHS model})
holds. Note first that: 
\begin{equation}
p\left(x_{b}|x_{a}\right)=\int d\lambda p\left(x_{b},\lambda|x_{a}\right),\label{Walborn Proof 1}
\end{equation}
with 
\begin{equation}
p\left(x_{b},\lambda|x_{a}\right)=p\left(\lambda|x_{a}\right)p\left(x_{b}|x_{a},\lambda\right)=p\left(\lambda|x_{a}\right)p_{\lambda}\left(x_{b}\right),\label{Walborn Proof 2}
\end{equation}
where the last equality holds since the state admits a LHS model.
Given $x_{a}$ we consider the relative entropy between $p\left(x_{b},\lambda|x_{a}\right)$
and $p\left(\lambda|x_{a}\right)p\left(x_{b}|x_{a}\right),$ which
is always nonnegative. Namely we have: 
\begin{equation}
\sum_{b}\int d\lambda p\left(x_{b},\lambda|x_{a}\right)\log_{2}\left(\frac{p\left(x_{b},\lambda|x_{a}\right)}{p\left(\lambda|x_{a}\right)p\left(x_{b}|x_{a}\right)}\right)\geq0.\label{Walborn Proof 3}
\end{equation}
The above can be written as a sum of two terms. The first is given
by:
\begin{align}
-\sum_{b}\int d\lambda p\left(x_{b},\lambda|x_{a}\right)\log_{2}\left(p\left(x_{b}|x_{a}\right)\right)=\label{Walborn Proof 4}\\
-\sum_{b}p\left(x_{b}|x_{a}\right)\log_{2}\left(p\left(x_{b}|x_{a}\right)\right)= & H\left(X_{B}|X_{A}=x_{a}\right).\nonumber 
\end{align}
The second, by using (\ref{Walborn Proof 2}), can be expressed as:
\begin{align}
\int d\lambda p\left(\lambda|x_{a}\right)\sum_{b}p_{\lambda}\left(x_{b}\right)\log_{2}\left(p_{\lambda}\left(x_{b}\right)\right)=\nonumber \\
-\int d\lambda p\left(\lambda|x_{a}\right)H_{\lambda}\left(X_{B}\right).\label{Walborn Proof 5}
\end{align}
Therefore (\ref{Walborn Proof 3}) implies: 
\begin{equation}
H\left(X_{B}|X_{A}=x_{a}\right)\geq\int d\lambda p\left(\lambda|x_{a}\right)H_{\lambda}\left(X_{B}\right),
\end{equation}
which leads to (\ref{Walborn ineq}) by averaging over $x_{a}$, that
provides a sufficient condition to detect steering states, indeed
any violation of it implies the presence of bipartite quantum steering.
If we now consider a sum as $\sum_{O=X,Z}H\left(O_{B}|O_{A}\right)$,
we find: 
\begin{equation}
\sum_{O=X,Z}H\left(O_{B}|O_{A}\right)\geq\int d\lambda q(\lambda)\sum_{O=X,Z}H_{\lambda}\left(O_{B}\right).\label{Bipartite EUR 1}
\end{equation}
In the right-hand side of (\ref{Bipartite EUR 1}) $\sum_{O=X,Z}H_{\lambda}\left(O_{B}\right)$
depends on $\lambda$, namely the two entropies are computed over
the state $\rho_{\lambda}^{B}$. However for any state Maaseen and
Uffink's EUR \cite{MUF} holds, namely we have: $\sum_{O=X,Z}H_{\lambda}\left(O_{B}\right)\geq\log_{2}\frac{1}{\alpha_{B}}$
, which together with $\int d\lambda q(\lambda)=1$, implies (\ref{MUF steering}):
\begin{equation}
H\left(X_{B}|X_{A}\right)+H\left(Z_{B}|Z_{A}\right)\geq-\log_{2}\alpha_{B}.
\end{equation}
Since the above must be valid for any nonsteerable state, any violation
of it indicates the presence of a steerable state. 

\section{Multipartite quantum steering}

In this section we start reviewing the concept of quantum steering
for multipartite systems. We focus on the tripartite case, where there
are two possible scenarios, following the approach given in \cite{Review,Multisteering1}.
In the first case, which can be named \emph{one-to-two steering scenario,}
Alice measures her system and wants to nonlocally influence the state
of the other two. The available information is encoded in the following
ensemble of unnormalized states: 
\begin{equation}
\sigma_{x_{a}}^{BC}=\Tr_{A}\left[\left(\Pi_{x_{a}}^{A}\otimes\mathbb{I}^{B}\otimes\mathbb{I}^{C}\right)\rho^{ABC}\right],\label{Multipartite Ensemble 12}
\end{equation}
where $\left\{ \Pi_{x_{a}}^{A}\right\} _{x_{a}}$ is a POVM of Alice's
measurements. The second possibility, the \emph{two-to-one steering
scenario, }consists in two parties, say Alice and Bob that, by measuring
their systems, want to influence the states of the third party. In
this case, the post-measured ensemble of states is given by: 
\begin{equation}
\sigma_{x_{a},x_{b}}^{C}=\Tr_{AB}\left[\left(\Pi_{x_{a}}^{A}\otimes\Pi_{x_{b}}^{B}\otimes\mathbb{I}^{C}\right)\rho^{ABC}\right],\label{Multipartite Esemble 21}
\end{equation}
where $\left\{ \Pi_{x_{a}}^{A}\right\} _{x_{a}}$,$\left\{ \Pi_{x_{b}}^{B}\right\} _{x_{b}}$
are POVMs of Alice and Bob's respectively. Multipartite steering scenario
therefore consists of all the asymmetric scenarios, where some subset
of the parties have full control on their subsystems, and they want
to steer the state of the remaining subsets. Just like entanglement,
which has a much richer structure in the multipartite case in than
the bipartite one since different notions of separability can be introduced,
also steerability have different levels for multipartite systems.
In the case of a tripartite system we have two notions: multipartite
steering and the genuine multipartite steering, which refers to the
impossibility to explain the correlations between measurement outcomes
in terms of different LHS models.

\section{one-to-two steering scenario}

\subsection{LHS models}

Let us first focus on the \emph{one-to-two steering scenario.} If
Alice cannot nonlocally influence Bob and Charlie the ensemble (\ref{Multipartite Ensemble 12})
becomes: 
\begin{equation}
\sigma_{x_{a}}^{BC}=\int d\lambda q(\lambda)p\left(x_{a}|\lambda\right)\rho_{\lambda}^{B}\otimes\rho_{\lambda}^{C}.\label{N Steering esemble}
\end{equation}
In the above there is no steering from Alice to Bob and Charlie and
each member of the ensemble is prepared in a separable state of Bob
and Charlie. Note that the above can be thought as a multipartite
LHS model where, with probabilities $q(\lambda)$ Alice receives $\lambda$
and outputs $x_{a}$ with probability $p\left(x_{a}|\lambda\right)$,
while Bob and Charlie's states are pre-determined by the value of
$\lambda.$ Any tripartite state that can produce an ensemble that
cannot be written as (\ref{N Steering esemble}) is said to be multipartite
steering. An example is provided by $\ket{\phi^{+}}\bra{\phi^{+}}^{AB}\otimes\rho^{C}$,
where $\ket{\phi^{+}}=\frac{1}{\sqrt{2}}\left(\ket{00}+\ket{11}\right)$.
Indeed Alice can prepare an ensemble that cannot be written as (\ref{N Steering esemble}).
This LHS model can be expressed in terms of joint probabilities as
(\ref{A-BC LHS non steering}), indeed: 
\begin{equation}
p\left(x_{a},x_{b},x_{c}\right)=\Tr_{BC}\left[\left(\Pi_{x_{b}}^{B}\otimes\Pi_{x_{c}}^{C}\right)\sigma_{x_{a}}^{BC}\right],\label{Prb}
\end{equation}
which implies (\ref{A-BC LHS non steering}) by using (\ref{N Steering esemble}).
Note that a slightly different definition of this form of multipartite
steering exists \cite{Multisteering1}. Indeed, one could require
that entanglement between Bob and Charlie is present. As a consequence,
Bob and Charlie's pre-determined state would be $\rho_{\lambda}^{BC}$,
instead of $\rho_{\lambda}^{B}\otimes\rho_{\lambda}^{C}$. However,
here we consider only the case where there is no entanglement between
Bob and Charlie, since we are interested in detecting the possible
simplest form of these quantum correlations. \\
If the state is non-GMS then the ensemble (\ref{Multipartite Ensemble 12})
can be expressed as:
\begin{align}
\sigma_{x_{a}}^{BC}= & \int_{A}d\nu q_{A}\left(\nu\right)p\left(x_{a}|\nu\right)\rho_{\nu}^{BC}+\int_{B}d\gamma q_{B}(\gamma)\rho_{\gamma}^{B}\otimes\sigma_{x_{a},\gamma}^{C}\nonumber \\
 & +\int_{C}d\omega q_{C}(\omega)\rho_{\omega}^{C}\otimes\sigma_{x_{a},\omega}^{B},\label{N GMS steering ensemble}
\end{align}
where $\sigma_{x_{a},\gamma}^{C}=\Tr_{A}\left[\left(\Pi_{x_{a}}^{A}\otimes\mathbb{I}^{C}\right)\rho_{\gamma}^{AC}\right]$
and $\sigma_{x_{a},\omega}^{B}=\Tr_{A}\left[\left(\Pi_{x_{a}}^{A}\otimes\mathbb{I}^{B}\right)\rho_{\omega}^{AB}\right]$
. Each member of this ensemble can be expressed as a sum of three
terms. In the first one there is no steering between Alice and Bob-Charlie.
In the other two, which are made of separable states only of Bob and
Charlie, Alice can steer one of the two subsystems but not the other.
This can be thought in terms of a hybrid-LHS model in the following
way. The hidden variable $\lambda$ discriminates between different
situations: in the first the global state of Bob and Charlie is pre-determined,
that is $\rho_{\lambda}^{BC}$, and this state can be entangled; in
the other two $\lambda$ determines the state of just one subsystem,
the other is not pre-determined. The previous example $\ket{\phi^{+}}\bra{\phi^{+}}^{AB}\otimes\rho^{C}$
can now lead to an ensemble of the form (\ref{N GMS steering ensemble}).
Any tripartite states that cannot produce an ensemble such (\ref{N GMS steering ensemble})
is said to be genuine multipartite steering. Conversely if (\ref{N GMS steering ensemble})
can be produced, the state is non-GMS. By using (\ref{Prb}) and (\ref{N GMS steering ensemble})
we can express this hybrid-LHS model in terms of joint probabilities:
\begin{align}
p\left(x_{a},x_{b},x_{c}\right)= & \varint_{A}d\nu q_{A}\left(\nu\right)p\left(x_{a}|\nu\right)p_{\nu}\left(x_{b},x_{c}\right)\label{A-BC NON-GMS-1}\\
 & +\varint_{B}d\gamma q_{B}\left(\gamma\right)p\left(x_{a}|\gamma\right)p_{\gamma}\left(x_{b}\right)p\left(x_{c}|x_{a},\gamma\right)\nonumber \\
 & +\varint_{C}d\omega q_{C}\left(\omega\right)p\left(x_{a}|\omega\right)p\left(x_{b}|x_{a},\omega\right)p_{\omega}\left(x_{c}\right),\nonumber 
\end{align}
which is exactly the requirement (\ref{A-BC NON-GMS}). \\

\subsection{Entropic uncertainty multipartite steering inequalities }

In this section we derive entropic steering inequalities for a multipartite
system, with the aim to discriminate also the different notions of
multipartite steering. We start by considering a nonsteerable state
and show that it must imply some inequalities, then we use them to
formulate sufficient conditions for multipartite steering detection.
If the state is nonsteerable it satisfies: 
\begin{equation}
p\left(x_{b},x_{c}|x_{a}\right)=\int d\lambda p\left(x_{b},x_{c},\lambda|x_{a}\right),\label{P1}
\end{equation}
with 
\begin{equation}
p\left(x_{b},x_{c},\lambda|x_{a}\right)=p\left(\lambda|x_{a}\right)p_{\lambda}\left(x_{b},x_{c}\right),\label{P2}
\end{equation}
where the last equality holds since (\ref{A-BC LHS non steering})
holds. As in the bipartite case, we now consider the relative entropy
between $p\left(x_{b},x_{c},\lambda|x_{a}\right)$ and $p\left(\lambda|x_{a}\right)p\left(x_{b},x_{c}|x_{a}\right)$,
which has to verify: 
\begin{equation}
\sum_{b,c}\int d\lambda p\left(x_{b},x_{c},\lambda|x_{a}\right)\log_{2}\left(\frac{p\left(x_{b},x_{c},\lambda|x_{a}\right)}{p\left(\lambda|x_{a}\right)p\left(x_{b},x_{c}|x_{a}\right)}\right)\geq0.
\end{equation}
The above, with (\ref{P1}) and (\ref{P2}), implies: 
\begin{equation}
H\left(X_{B},X_{C}|X_{A}=x_{a}\right)\geq\int d\lambda p\left(\lambda|x_{a}\right)H_{\lambda}\left(X_{B},X_{C}\right);
\end{equation}
and by averaging over $x_{a}$ we arrive at:
\begin{equation}
H\left(X_{B},X_{C}|X_{A}\right)\geq\int d\lambda q\left(\lambda\right)H_{\lambda}\left(X_{B},X_{C}\right).
\end{equation}
Since Bob and Charlie share a separable state, for $m=B,C$ we also
have: 
\begin{equation}
p\left(x_{m},\lambda|x_{a}\right)=p\left(\lambda|x_{a}\right)p_{\lambda}\left(x_{m}\right).\label{Pr}
\end{equation}
Now by considering the relative entropy between $p\left(x_{m},\lambda|x_{a}\right)$
and $p\left(\lambda|x_{a}\right)p\left(x_{m}|x_{a}\right)$ for $m=B,C$,
we can derive in the same way: 
\begin{equation}
H\left(X_{m}|X_{A}\right)\geq\int d\lambda q\left(\lambda\right)H_{\lambda}\left(X_{m}\right).\label{P=0000A3}
\end{equation}
The entropic uncertainty relations (\ref{A-BC NS EUR1}), namely $\sum_{O=X,Z}H\left(O_{m}|O_{A}O_{\bar{m}}\right)\geq-\log_{2}\alpha_{m}$,
can be derived simply by noting that the following holds for any state:
\begin{equation}
\sum_{O=X,Z}H_{\lambda}\left(O_{m}\right)\geq-\log_{2}\alpha_{m},\label{MUF""}
\end{equation}
with $m=B,C,BC$, since $\int d\lambda q(\lambda)=1.$ From the above
we can see that if we considered EUR different from the ones of \cite{MUF},
we could find other EUR for nonsteerable states. These relations can
be used to define sufficient conditions for multipartite steering.
Indeed any violation indicates its presence. \\
Let us consider complementary observables, namely observables whose
eigenbasis are mutually unbiased \footnote{We remind that two basis $\left\{ \ket{a_{i}}\right\} _{i=0}^{d-1}$
, $\left\{ \ket{b_{j}}\right\} _{j=0}^{d-1}$ are said to be mutually
unbiased if $\left|\braket{a_{i}|b_{j}}\right|=\frac{1}{\sqrt{d}}$
for all $j$ and k, where $d$ is the dimension of the system. For
more details we refer to \cite{MUBs,MUBs2}.}. In this case we have: 
\begin{equation}
\sum_{O=X,Z}H_{\lambda}\left(O_{BC}\right)\geq2\log_{2}d_{BC};\label{EUR MUbs}
\end{equation}
\begin{equation}
\sum_{O=X,Z}H_{\lambda}\left(O_{m}\right)\geq\log_{2}d_{m},\label{EUR MUBs 2}
\end{equation}
for $m=B,C$; being $d_{m}$ the dimension of the system $m$ and
$d_{BC}=d_{B}d_{C}.$ \\
We want now to extend these results to the case of GMS states. Suppose
now that the state of the system is non-GMS, then we have: 
\begin{align}
p\left(x_{b},x_{c}|x_{a}\right) & =\int_{A}d\nu p\left(x_{b},x_{c},\nu|x_{a}\right)+\int_{B}d\gamma p\left(x_{b},x_{c},\gamma|x_{a}\right)\nonumber \\
 & +\int_{C}d\omega p\left(x_{b},x_{c},\omega|x_{a}\right).\label{PR1}
\end{align}
Since (\ref{A-BC NON-GMS-1}) holds, each term can be written as follows:
\begin{equation}
p\left(x_{b},x_{c},\nu|x_{a}\right)=p\left(\nu|x_{a}\right)p_{\nu}\left(x_{b},x_{c}\right);\label{PR2}
\end{equation}
\begin{equation}
p\left(x_{b},x_{c},\gamma|x_{a}\right)=p\left(\gamma|x_{a}\right)p_{\gamma}\left(x_{b}\right)p\left(x_{c}|x_{a},\gamma\right);\label{PR3}
\end{equation}
\begin{equation}
p\left(x_{b},x_{c},\omega|x_{a}\right)=p\left(\omega|x_{a}\right)p_{\omega}\left(x_{c}\right)p\left(x_{b}|x_{a},\omega\right).\label{PR4}
\end{equation}
Equation \ref{PR1} can be also expressed as $p\left(x_{b},x_{c}|x_{a}\right)=\int d\lambda p\left(x_{b},x_{c},\lambda|x_{a}\right)$
where $\lambda$ is a classical variable such that $\int d\lambda q(\lambda)=\int_{A}d\nu q_{A}(\nu)+\int_{B}d\gamma q_{B}(\gamma)+\int_{C}d\omega q_{C}(\omega)=1.$
We consider now the relative entropy between $p\left(x_{b},x_{c},\lambda|x_{a}\right)$
and $p\left(\lambda|x_{a}\right)p\left(x_{b},x_{c}|x_{a}\right),$
which must be nonnegative: 
\begin{equation}
\sum_{b,c}\int d\lambda p\left(x_{b},x_{c},\lambda|x_{a}\right)\log_{2}\left(\frac{p\left(x_{b},x_{c},\lambda|x_{a}\right)}{p\left(\lambda|x_{a}\right)p\left(x_{b},x_{c}|x_{a}\right)}\right)\geq0.
\end{equation}
The above quantity is a sum of two terms. The first one is: 
\begin{equation}
-\sum_{b,c}\int d\lambda p\left(x_{b},x_{c},\lambda|x_{a}\right)\log_{2}p\left(x_{b},x_{c}|x_{a}\right)
\end{equation}
that is $H\left(X_{B},X_{C}|X_{A}=X_{a}\right),$ since (\ref{PR1})
holds. The second one is: 
\begin{equation}
\sum_{b,c}\int d\lambda p\left(x_{b},x_{c},\lambda|x_{a}\right)\log_{2}\left(\frac{p\left(x_{b},x_{c},\lambda|x_{a}\right)}{p\left(\lambda|x_{a}\right)}\right),\label{ssshs}
\end{equation}
which, by using the decomposition of $p\left(x_{b},x_{c},\lambda|x_{a}\right)$
given by (\ref{PR2}), (\ref{PR3}) and (\ref{PR4}), can be written
as a sum of three terms: 
\begin{align}
-\int_{A}d\nu p\left(\nu|x_{a}\right)H_{\nu}\left(X_{b},X_{c}\right);
\end{align}
\begin{equation}
-\int_{B}d\gamma p\left(\gamma|x_{a}\right)\left[H_{\gamma}\left(X_{b}\right)+H\left(X_{c}|X_{A}=x_{a},\gamma\right)\right];
\end{equation}
\begin{equation}
-\int_{C}d\lambda p\left(\omega|x_{a}\right)\left[H_{\omega}\left(X_{c}\right)+H\left(X_{b}|X_{A}=x_{a},\omega\right)\right].
\end{equation}
After reordering the terms and averaging over $x_{a}$, that for example
implies $\sum_{a}p\left(x_{a}\right)p\left(\gamma|x_{a}\right)=q_{B}\left(\gamma\right)$
and similar relations, we arrive at: 
\begin{flalign}
H\left(X_{BC}|X_{A}\right) & \geq\int_{A}d\nu q_{A}\left(\nu\right)H_{\nu}\left(X_{BC}\right)\label{A-BC NGMS Walborn A-1}\\
 & +\int_{B}d\gamma q_{B}\left(\gamma\right)H_{\gamma}\left(X_{B}\right)\nonumber \\
 & +\int_{C}d\omega q_{C}\left(\omega\right)H_{\omega}\left(X_{C}\right)\nonumber \\
 & +\int_{B}d\gamma q_{B}\left(\gamma\right)H\left(X_{C}|X_{A},\gamma\right)\nonumber \\
 & +\int_{C}d\omega q_{C}\left(\omega\right)H\left(X_{B}|X_{A},\omega\right),\nonumber 
\end{flalign}
where $X_{BC}=X_{B}\otimes X_{C}.$ Since $H\left(X_{C}|X_{A},\gamma\right)\geq0$
and $H\left(X_{B}|X_{A},\omega\right)\geq0$ for any $\gamma$ and
$\omega,$ we finally arrive at: 
\begin{align}
H\left(X_{BC}|X_{A}\right) & \geq\int_{A}d\nu q_{A}\left(\nu\right)H_{\nu}\left(X_{BC}\right)\label{NON GMS W1}\\
 & +\int_{B}d\gamma q_{B}\left(\gamma\right)H_{\gamma}\left(X_{B}\right)\nonumber \\
 & +\int_{C}d\omega q_{C}\left(\omega\right)H_{\omega}\left(X_{C}\right).\nonumber 
\end{align}
Then by using (\ref{MUF""}) we can derive the following state-dependent
entropic uncertainty relations: 
\begin{align}
\sum_{O=X,Z}H\left(O_{BC}|O_{A}\right)\geq & -\int_{A}d\nu q_{A}\left(\nu\right)\log_{2}\alpha_{BC}\label{NON GMS EUR!}\\
 & -\int_{B}d\gamma q_{B}\left(\gamma\right)\log_{2}\alpha_{B}\nonumber \\
 & -\int_{C}d\omega q_{C}\left(\omega\right)\log_{2}\alpha_{C}.\nonumber 
\end{align}
Note that in general $\alpha_{BC}\geq\min\left\{ \alpha_{B},\alpha_{C}\right\} =\alpha_{min},$
hence the above implies: 
\begin{equation}
\sum_{O=X,Z}H\left(O_{BC}|O_{A}\right)\geq-\log_{2}\alpha_{min},\label{AA}
\end{equation}
which for complementary observables becomes:
\begin{equation}
\sum_{O=X,Z}H\left(O_{BC}|O_{A}\right)\geq\log_{2}d_{\min},\label{AB}
\end{equation}
where $d_{\min}=\min\left\{ d_{B},d_{C}\right\} $. \\
Now we focus on the conditional entropies $H\left(X_{B}|X_{A}\right)$
and $H\left(X_{C}|X_{A}\right)$. Since $p\left(x_{b}|x_{a}\right)=\sum_{c}p\left(x_{b},x_{c}|x_{a}\right)$
the three terms (\ref{PR2}), (\ref{PR3}) and (\ref{PR4}) become:
\begin{equation}
p\left(x_{b},\nu|x_{a}\right)=p\left(\nu|x_{a}\right)p_{\nu}\left(x_{b}\right);\label{PR2-1}
\end{equation}
\begin{equation}
p\left(x_{b},\gamma|x_{a}\right)=p\left(\gamma|x_{a}\right)p_{\gamma}\left(x_{b}\right);\label{PR3-1}
\end{equation}
\begin{equation}
p\left(x_{b},\omega|x_{a}\right)=p\left(\omega|x_{a}\right)p\left(x_{b}|x_{a},\omega\right).\label{PR4-1}
\end{equation}
From the above relations we derive, with the same arguments that we
have used in the previous case, 
\begin{flalign}
H\left(X_{B}|X_{A}\right) & \geq\int_{A}d\nu q_{A}\left(\nu\right)H_{\nu}\left(X_{B}\right)\label{A-BC NGMS Walborn A-1-1}\\
 & +\int_{B}d\gamma q_{B}\left(\gamma\right)H_{\gamma}\left(X_{B}\right)\nonumber \\
 & +\int_{C}d\omega q_{C}\left(\omega\right)H\left(X_{B}|X_{A},\omega\right).\nonumber 
\end{flalign}
The same holds for Charlie: 
\begin{align}
H\left(X_{C}|X_{A}\right) & \geq\int_{A}d\nu q_{A}\left(\nu\right)H_{\nu}\left(X_{C}\right)\label{CHarlie W}\\
 & +\int_{C}d\omega q_{C}\left(\omega\right)H_{\omega}\left(X_{C}\right)\nonumber \\
 & +\int_{B}d\omega q_{B}\left(\gamma\right)H\left(X_{C}|X_{A},\gamma\right).\nonumber 
\end{align}
Any violation of one of the above EUR indicates that the state is
GMS. In order to define stronger steering criteria we can also look
at conditional entropies of the form $H\left(X_{B}|X_{A},X_{C}\right)$
and $H\left(X_{C}|X_{A},X_{B}\right)$ where measurements on parts
different from Alice are performed. As an example we consider $H\left(X_{C}|X_{A},X_{B}\right),$
therefore we are interested in the probability:
\begin{align}
p\left(x_{c}|x_{a},x_{b}\right) & =\int_{A}d\nu p\left(x_{c},\nu|x_{a},x_{b}\right)+\int_{B}d\gamma p\left(x_{c},\gamma|x_{a},x_{b}\right)\nonumber \\
 & +\int_{C}d\omega p\left(x_{c},\omega|x_{a},x_{b}\right).
\end{align}
Since the state is non-GMS, the terms in the above equation can be
written as follows:
\begin{equation}
p\left(x_{c},\nu|x_{a},x_{b}\right)=p\left(\nu|x_{a},x_{b}\right)p\left(x_{c}|x_{b},\nu\right);
\end{equation}
\begin{equation}
p\left(x_{c},\gamma|x_{a},x_{b}\right)=p\left(\gamma|x_{a},x_{b}\right)p\left(x_{c}|x_{a},\gamma\right);
\end{equation}
\begin{equation}
p\left(x_{c},\lambda|x_{a},x_{b}\right)=p\left(\omega|x_{a},x_{b}\right)p_{\omega}\left(x_{c}\right).
\end{equation}
From the above we can derive in the usual way that:
\begin{equation}
H\left(X_{C}|X_{A},X_{B}\right)\geq\int_{C}d\omega q_{C}\left(\omega\right)H_{\omega}\left(X_{C}\right).
\end{equation}
The same holds for Bob:
\begin{equation}
H\left(X_{B}|X_{A},X_{C}\right)\geq\int_{B}d\gamma q_{B}\left(\gamma\right)H_{\gamma}\left(X_{B}\right).
\end{equation}
In terms of entropic uncertainty relations we have:
\begin{equation}
\sum_{O=X;Z}H\left(O_{B}|O_{A},O_{C}\right)\geq-\int_{B}d\gamma q_{B}\left(\gamma\right)\log_{2}\alpha_{B},
\end{equation}
\begin{equation}
\sum_{O=X;Z}H\left(O_{C}|O_{A},O_{B}\right)\geq-\int_{C}d\omega q_{C}\left(\omega\right)\log_{2}\alpha_{C}.
\end{equation}
these two equations are equivalent to eqs. (\ref{eq 9}). Moreover,
when combined with (\ref{NON GMS EUR!}) they imply eq. (\ref{EUR Fin}),
namely: 
\begin{equation}
\sum_{O=X,Z}A\left(O_{A},O_{B},O_{C}\right)\geq-2\log_{2}\alpha_{min},\label{HH}
\end{equation}
where $A\left(O_{A},O_{B},O_{C}\right)=H\left(O_{BC}|O_{A}\right)+\sum_{m=B,C}H\left(O_{m}|O_{A},O_{\bar{m}}\right)$
and $\alpha_{min}=\min\left\{ \alpha_{B},\alpha_{C}\right\} $ \\
A nonsteerable state satisfies equation (\ref{33-1}) instead, namely:
\begin{equation}
\sum_{O=X,Z}A\left(O_{A},O_{B},O_{C}\right)\geq-2\log_{2}\alpha_{BC}\geq-4\log_{2}\alpha_{min}.\label{33}
\end{equation}
Indeed for a nonsteerable state we have shown that $\sum_{O=X,Z}H\left(O_{BC}|O_{A}\right)\geq-\log_{2}\alpha_{BC}$.
Then in this case we also have: 
\begin{equation}
p\left(x_{c}|x_{a},x_{b}\right)=\int d\lambda p\left(x_{c},\lambda|x_{a},x_{b}\right),
\end{equation}
with $p\left(x_{c},\lambda|x_{a},x_{b}\right)=p\left(\lambda|x_{a},x_{b}\right)p_{\lambda}\left(x_{c}\right)$,
which by using the usual procedure leads to:
\begin{equation}
\sum_{O=X,Z}H\left(O_{C}|O_{A}O_{B}\right)\geq-\log_{2}\alpha_{C}.
\end{equation}
The same holds for Bob:
\begin{equation}
\sum_{O=X,Z}H\left(O_{C}|O_{A}O_{B}\right)\geq-\log_{2}\alpha_{B}.
\end{equation}
The above equations, together $\sum_{O=X,Z}H\left(O_{BC}|O_{A}\right)\geq-\log_{2}\alpha_{BC},$
lead to (\ref{33}), which is expressed in terms of $\alpha_{min}$.
\\
As a final result we want to derive a state-dependent entropic uncertainty
relation starting from (\ref{A-BC NGMS Walborn A-1}), which we remind
that it is valid for any non-GMS states. To derive (\ref{NON GMS W1})
from (\ref{A-BC NGMS Walborn A-1}) we have used the fact that the
conditional entropies must be always greater than zero, however we
can take advantage of these terms by considering the entropic uncertainty
relations in presence of quantum memories \cite{Berta1,Haseli,Haseli2,key-17,JS3},
which, for the cases considered, can be expressed in terms of conditional
Shannon entropies \cite{JS3} as:
\begin{equation}
\sum_{o=X,Z}H\left(O_{C}|O_{A},\gamma\right)\geq-\log_{2}\alpha_{C}+S_{\gamma}\left(C|A\right),\label{Quantum memory CA}
\end{equation}
\begin{equation}
\sum_{o=X,Z}H\left(O_{B}|O_{A},\omega\right)\geq-\log_{2}\alpha_{B}+S_{\omega}\left(B|A\right),\label{Quantum memory BA}
\end{equation}
where $S_{\gamma}\left(C|A\right)$ represents the Von Neumann conditional
entropy between Charlie and Alice over the state $\rho_{\gamma}^{AC}$
and $S_{\omega}\left(B|A\right)$ the Von Neumann conditional entropy
between Bob and Alice over $\rho_{\omega}^{AB}.$ Note that these
quantities can be negative \cite{Cerf} for entangled states, moreover
their lowest values $-\log_{2}d_{C}$ and $-\log_{2}d_{B}$ are reached
by maximally entangled states.\\
Therefore, by using (\ref{MUF""}), (\ref{Quantum memory CA}) and
(\ref{Quantum memory BA}), the quantity $\sum_{O=X,Z}H\left(O_{BC}|O_{A}\right)$
can be lower bounded as follows: 
\begin{gather}
\sum_{O=X,Z}H\left(O_{BC}|O_{A}\right)\geq-\log_{2}\alpha_{CB}\nonumber \\
+\int_{B}d\gamma q_{B}\left(\gamma\right)S_{\gamma}\left(C|A\right)+\int_{C}d\omega q_{C}\left(\omega\right)S_{\omega}\left(B|A\right),\label{memory eurP}
\end{gather}
where we have used also the relation $\log_{2}\alpha_{BC}=\log_{2}\alpha_{B}+\log_{2}\alpha_{C}.$
Then we can conclude that for a non-GMS state the following inequality
holds (\ref{memory eurP-1}):
\begin{gather}
\sum_{O=X,Z}H\left(O_{BC}|O_{A}\right)\geq-\log_{2}\alpha_{CB}\nonumber \\
+\int_{B}d\gamma q_{B}\left(\gamma\right)S_{\gamma}\left(C|A\right)+\int_{C}d\omega q_{C}\left(\omega\right)S_{\omega}\left(B|A\right).\label{memory eurP-1-1}
\end{gather}

\section{Two-to-one steering scenario}

\subsection{LHS models}

We now consider the \emph{two-to-one steering scenario, }where Alice
and Bob want to nonlocally influence Charlie's state. In the case
of a nonsteerable state the ensemble (\ref{Multipartite Esemble 21})
becomes:
\begin{equation}
\sigma_{x_{a},x_{b}}^{C}=\int d\lambda q(\lambda)p\left(x_{a}|\lambda\right)p\left(x_{b}|\lambda\right)\rho_{\lambda}^{C},\label{NONSteerable 2-1}
\end{equation}
which can be written in term of probabilities as: 
\begin{equation}
p\left(x_{a},x_{b},x_{c}\right)=\int d\lambda q(\lambda)p\left(x_{a}|\lambda\right)p\left(x_{b}|\lambda\right)p_{\lambda}\left(x_{c}\right).\label{Non steerable 2-1 prob}
\end{equation}
In the above Alice and Bob can share local correlations but are jointly
unable to steer Charlie. This case corresponds to a LHS model where,
conversely from the previous scenario Bob, receives $\lambda$ instead
of $\rho_{\lambda}^{B}$ and obtains $x_{b}$ with probability $p\left(x_{b}|\lambda\right).$
Charlie, as in the previous, is represented by a pre-determined state
$\rho_{\lambda}^{C}$ which depends solely on $\lambda$. If the ensemble
(\ref{Multipartite Esemble 21}) cannot be written as (\ref{NONSteerable 2-1})
we say that the state is steerable from Alice and Bob to Charlie.
For a non-GMS state the ensemble (\ref{Multipartite Esemble 21})
becomes:
\begin{align}
\sigma_{x_{a},x_{b}}^{C} & =\varint_{A}d\nu q_{A}\left(\nu\right)p\left(x_{a}|\nu\right)\sigma_{x_{b},\nu}^{C}\nonumber \\
 & +\varint_{B}d\gamma q_{B}\left(\gamma\right)p\left(x_{b}|\gamma\right)\sigma_{x_{a},\gamma}^{C}\nonumber \\
 & +\varint_{C}d\omega q_{AB}\left(\omega\right)p\left(x_{a},x_{b}|\omega\right)\rho_{\omega}^{C},\label{AB-C NON-GMS-2}
\end{align}
which is as a sum of three terms; in the first only Bob can steer
Charlie, in the second only Alice can steer the state of Charlie,
whereas in the third Alice and Bob cannot jointly steer Charlie, but
they can share quantum correlations. In terms of probabilities we
have: 

\begin{align}
p\left(x_{a},x_{b},x_{c}\right) & =\varint_{A}d\nu q_{A}\left(\nu\right)p\left(x_{a}|\nu\right)p\left(x_{b}|\nu\right)p\left(x_{c}|x_{b}\nu\right)\nonumber \\
 & +\varint_{B}d\gamma q_{B}\left(\gamma\right)p\left(x_{a}|\gamma\right)p\left(x_{b}|\gamma\right)p\left(x_{c}|x_{a},\gamma\right)\nonumber \\
 & +\varint_{C}d\omega q_{C}\left(\omega\right)p\left(x_{a},x_{b}|\omega\right)p_{\omega}\left(x_{c}\right).\label{AB-C NON-GMS-1}
\end{align}

\subsection{Entropic steering inequalities}

Let us start with a nonsteerable state: it is characterized by: 
\begin{equation}
p\left(x_{c}|x_{a},x_{b}\right)=\int d\lambda p\left(x_{c},\lambda|x_{a},x_{b}\right),\label{2-1 NS}
\end{equation}
with 
\begin{equation}
p\left(x_{c},\lambda|x_{a},x_{b}\right)=p\left(\lambda|x_{a},x_{b}\right)p_{\lambda}\left(x_{c}\right).\label{2-1 NS2}
\end{equation}
The above relation allows us to consider, given $x_{a}$ and $x_{b}$,
the relative entropy between $p\left(x_{c},\lambda|x_{a},x_{b}\right)$
and $p\left(x_{c}|x_{a},x_{b}\right)p\left(\lambda|x_{a},x_{b}\right)$
that by definition is always nonnegative, namely we have: {\small{}
\begin{equation}
\sum_{c}\int d\lambda p\left(x_{c},\lambda|x_{a},x_{b}\right)\log_{2}\left(\frac{p\left(x_{c},\lambda|x_{a},x_{b}\right)}{p\left(\lambda|x_{a},x_{b}\right)p\left(x_{c}|x_{a},x_{b}\right)}\right)\geq0.\label{2-1 NS3}
\end{equation}
}By proceeding as in the previous section we can arrive at: 
\begin{equation}
H\left(X_{C}|X_{A},X_{B}\right)\geq\int dq(\lambda)H_{\lambda}\left(X_{C}\right),\label{2-1 Walborn}
\end{equation}
where also in this case $\lambda$ indicates that the Shannon entropy
$H_{\lambda}\left(X_{C}\right)$ is calculated over the state $\rho_{\lambda}^{C}$.
By using the EUR $H_{\lambda}\left(X_{C}\right)+H_{\lambda}\left(Z_{C}\right)\geq-\log_{2}\alpha_{C}$
we arrive at: 
\begin{equation}
\sum_{O=X,Z}H\left(O_{C}|O_{A},O_{B}\right)\geq-\log_{2}\alpha_{C}.\label{EUR2-1 NS}
\end{equation}
For a nonsteerable state also the following holds: 
\begin{equation}
p\left(x_{c},\lambda|x_{m}\right)=p\left(\lambda|x_{m}\right)p_{\lambda}\left(x_{c}\right),\label{2-1 NS4}
\end{equation}
with $m=A,B$. With the same derivation for $m=A,B$ we can therefore
arrive at: 
\begin{equation}
H\left(X_{C}|X_{m}\right)\geq\int dq(\lambda)H_{\lambda}\left(X_{C}\right),\label{2-1 Walborn2}
\end{equation}
which then implies: 
\begin{equation}
\sum_{O=X,Z}H\left(O_{C}|O_{m}\right)\geq-\log_{2}\alpha_{C}.\label{EUR2-1 NS Marg}
\end{equation}
The entropic uncertainty relations (\ref{EUR2-1 NS}), (\ref{EUR2-1 NS Marg})
represent sufficient criteria to steerability from Alice and Bob to
Charlie. Note that if we sum the above inequality over $m$ we derive:
\begin{equation}
\sum_{m=A,B}\left(\sum_{O=X,Z}H\left(O_{C}|O_{m}\right)\right)\geq-2\log_{2}\alpha_{C}.\label{2-1 EUR FINALe}
\end{equation}
 Let us now focus on non-GMS states. The probability $p\left(x_{c}|x_{a},x_{b}\right)$
can be now written as: 
\begin{align}
p\left(x_{c}|x_{a},x_{b}\right) & =\int_{A}d\nu p\left(x_{c},\nu|x_{a},x_{b}\right)+\int_{B}d\gamma p\left(x_{c},\gamma|x_{a},x_{b}\right)\nonumber \\
 & +\int_{C}d\omega p\left(x_{c},\omega|x_{a},x_{b}\right).
\end{align}
The three terms in the above equation can be expressed as follow:
\begin{equation}
p\left(x_{c},\nu|x_{a},x_{b}\right)=p\left(\nu|x_{a},x_{b}\right)p_{\nu}\left(x_{c}|x_{b}\right);\label{2-1 PRA}
\end{equation}
\begin{equation}
p\left(x_{c},\gamma|x_{a},x_{b}\right)=p\left(\gamma|x_{a},x_{b}\right)p_{\gamma}\left(x_{c}|x_{a}\right);\label{2-1 PRB}
\end{equation}
\begin{equation}
p\left(x_{c},\omega|x_{a},x_{b}\right)=p\left(\omega|x_{a},x_{b}\right)p_{\omega}\left(x_{c}\right).\label{2-1 PRAB}
\end{equation}
Now we can consider the relative entropy between $p\left(x_{c},\lambda|x_{a},x_{b}\right)$
and $p\left(x_{c}|x_{a},x_{b}\right)p\left(\lambda|x_{a},x_{b}\right),$
where $\lambda$ is such that $\int d\lambda q(\lambda)=\int_{A}d\nu q_{A}(\nu)+\int_{B}d\gamma q_{B}(\gamma)+\int_{C}d\omega q_{C}(\omega)=1.$
The non-negativity of the relative entropy implies:{\small{}
\begin{equation}
\sum_{c}\int d\lambda p\left(x_{c},\lambda|x_{a},x_{b}\right)\log_{2}\left(\frac{p\left(x_{c},\lambda|x_{a},x_{b}\right)}{p\left(\lambda|x_{a},x_{b}\right)p\left(x_{c}|x_{a},x_{b}\right)}\right)\geq0.
\end{equation}
} By proceeding as in the derivation of (\ref{NON GMS W1}) we arrive
in this case at: 
\begin{equation}
H\left(X_{C}|X_{A},X_{B}\right)\geq\int_{C}d\omega q_{C}\left(\omega\right)H_{\omega}\left(X_{C}\right).\label{EUR 2-1 Walborn NONgms}
\end{equation}
We can also consider the relative entropy between $p\left(x_{c},\lambda|x_{a}\right)$
and $p\left(x_{c}|x_{a}\right)p\left(\lambda|x_{a}\right)$. The same
arguments used to derive (\ref{CHarlie W}) leads us to: 
\begin{equation}
H\left(X_{C}|X_{A}\right)\geq\int_{A}d\nu q_{A}(\nu)H_{\nu}\left(X_{C}\right)+\int_{C}d\omega q_{\omega}(\omega)H_{\omega}\left(X_{C}\right).\label{2-1 GMS Wal}
\end{equation}
The same holds by conditioning on Bob: 
\begin{equation}
H\left(X_{C}|X_{A}\right)\geq\int_{B}d\gamma q_{\gamma}(\gamma)H_{\gamma}\left(X_{C}\right)+\int_{C}d\omega q_{\omega}(\omega)H_{\omega}\left(X_{C}\right).\label{GMS Walb2}
\end{equation}
The corresponding EUR of (\ref{EUR 2-1 Walborn NONgms}, \ref{GMS Walb2})
are then given by: 
\begin{equation}
\sum_{O=X,Y}H\left(O_{C}|O_{A},O_{B}\right)\geq-\log_{2}\alpha_{C}\int_{C}d\omega q_{C}\left(\omega\right);
\end{equation}

{\small{}
\begin{equation}
\sum_{O=X,Y}H\left(O_{C}|O_{A}\right)\geq-\log_{2}\alpha_{C}\left(\int_{A}d\nu q_{A}(\nu)+\int_{C}d\omega q_{C}(\omega)\right);
\end{equation}
\begin{equation}
\sum_{O=X,Y}H\left(O_{C}|O_{B}\right)\geq-\log_{2}\alpha_{C}\left(\int_{B}d\gamma q_{B}(\gamma)+\int_{C}d\omega q_{C}(\omega)\right)
\end{equation}
}The sum of the two last inequalities above implies: 
\begin{equation}
\begin{array}{c}
\sum_{O=X,Y}\left(H\left(O_{C}|O_{A}\right)+H\left(O_{C}|O_{B}\right)\right)\geq\\
-\log_{2}\alpha_{C}\left(1+\int_{C}d\omega q_{C}(\omega)\right),
\end{array}
\end{equation}
then by using $-\log_{2}\alpha_{C}\int_{C}d\omega q_{C}(\omega)\geq0$,
we arrive at: 
\begin{equation}
\sum_{O=X,Y}\left(H\left(O_{C}|O_{A}\right)+H\left(O_{C}|O_{B}\right)\right)\geq-\log_{2}\alpha_{C},
\end{equation}
 which in the case of complementary observables is: 
\begin{equation}
\sum_{O=X,Y}\left(H\left(O_{C}|O_{A}\right)+H\left(O_{C}|O_{B}\right)\right)\geq\log_{2}d_{C}.
\end{equation}
 Any violation of the above inequality indicates the presence of genuine
multipartite steering from Alice and Bob to Charlie. 

\section{Steering detection}

In this section we study the steering detection power of the relations
derived in the previous sections. Here we consider only multiqubit
systems and complementary observables, namely $X$ and $Z$ are always
the Pauli matrices $\sigma_{x}$ and $\sigma_{z}$ respectively. We
study the following quantities: 
\begin{equation}
S_{1}=\sum_{O=X,Z}H\left(O_{BC}|O_{A}\right);
\end{equation}
\begin{equation}
S_{2}=\sum_{O=X,Z}H\left(O_{B}|O_{A}\right);
\end{equation}
\begin{equation}
S_{3}=\sum_{O=X,Z}H\left(O_{C}|O_{A}\right);
\end{equation}
\begin{equation}
C=\sum_{O=X,Z}H\left(O_{B}|O_{A},O_{C}\right);
\end{equation}
\begin{equation}
A=\sum_{O}H\left(O_{BC}|O_{A}\right)+\sum_{m=B,C}H\left(O_{m}|O_{A},O_{\bar{m}}\right).
\end{equation}
As a first example we consider the $GHZ$ class of three qubit states,
which can be expressed as: 
\begin{equation}
\ket{GHZ}=a\ket{000}+\sqrt{1-a^{2}}\ket{111},\label{GHZ}
\end{equation}
with $0\leq a\leq1$. One can check that $S_{2}$ and $S_{3}$ are
always greater or equal to $1$, that is the threshold below which
quantum steering is detected. Hence for $GHZ$ these two relations
do not see any form of steering. In Figure 1 
\begin{figure}
\caption{\protect\includegraphics[scale=0.37]{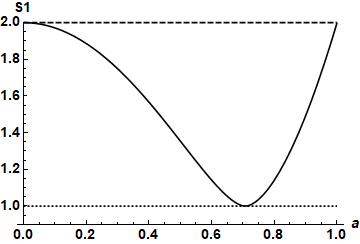}\protect \\
{\footnotesize{}Plot of $S_{1}$ for the state $a\ket{000}+\sqrt{1-a^{2}}\ket{111}$
as a function of $a.$ The upper dashed line represents the threshold
to detect steering (which is then seen for all $a$), the lower dashed
line is the threshold for GMS. }}
 
\end{figure}
 we plot $S_{1}$. As we can see for all $a$ this quantity is lower
than $2$, except for $a=0,1$, and therefore it detects quantum steering.
However since $S_{1}$ is always greater than $1$, we cannot detect
genuine tripartite steering. Also $C$ can detect all multipartite
steerable states for all $a\neq0,1$. A different result can be achieved
by considering the quantity $A,$ that is plotted in Figure 2 
\begin{figure}
\raggedright{}\caption{\protect\includegraphics[scale=0.37]{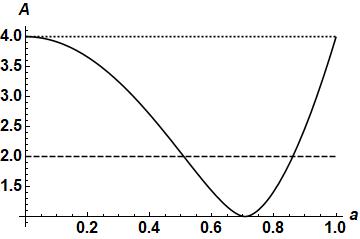}}
{\footnotesize{}Plot of $A$ for the state $a\ket{000}+\sqrt{1-a^{2}}\ket{111}$
as a function of $a.$ The upper dashed line represents the threshold
to detect steering (which is then seen for all $a$), the lower dashed
line is the threshold for GMS. Some GMS states are detected with $A.$}
\end{figure}
 for this class of states, indeed there exist states that violate
the inequality $A\geq2$. All the other states violate the inequality
$A_{4}\geq4$, hence they are all detected as tripartite steerable
states. \\
In order to compare the power in detecting steerability of the criteria
represented by $S_{1},C$ and $A$ we consider the robustness of steerability
of the standard $GHZ$, i.e. $a=\frac{1}{\sqrt{2}}$, under white
noise The state is represented as: 
\begin{equation}
\rho=p\ket{GHZ}_{S}\bra{GHZ}_{S}+\frac{1-p}{8}\mathbb{I},
\end{equation}
with $0\leq p\leq1$ and $\ket{GHZ}_{S}=\frac{1}{\sqrt{2}}\left(\ket{000}+\ket{111}\right)$.
The results for $A$ are shown in Figure 3. As we can see only for
$p>0.80$ the state is detected as steerable and only for few values
of $p$, namely for $p>0.95,$ is detected as genuine tripartite steerable.
The criterion $S_{1}<2$ detects the state as steerable for $p>0.86,$
while $C<1$ when $p>0.74.$ Therefore for this type of steering the
criterion related to $C$ outperforms the others. We also consider
the steerability robustnees of the standard $W$ state, that is $W=\frac{1}{\sqrt{3}}\left(\ket{001}+\ket{010}+\ket{100}\right)$,
under white noise, namely we consider the state:
\begin{equation}
\rho=p\ket{W}\bra{W}+\frac{1-p}{8}\mathbb{I}.
\end{equation}
 By considering $S_{1}$ this state is identified as multipartite
steerable for any $p>0.98$, while by using $C$ and $A$ we find
respectively $p>0.85$ and $p>0.91$. As in the previous case the
criterion based on $C$ detects more steerable states. However one
can check that $W$ does not violate any of our genuine multipartite
criteria. 

\section{Conclusions}

In conclusion, we derived and characterized a certain number of entropic
uncertainty inequalities whose violation guarantees the presence of
different classes of multipartite steering. Most of all these criteria
enable to distinguish between multipartite steering and genuine multipartite
steering and, being state-independent, they allow to study the steering
property of an unknown multipartite state. Recently, for bipartite
systems some new steering criteria based on generalized entropic uncertainty
relations, namely defined in terms of Tsallis entropies instead of
Shannon one, have been derived \cite{Tsallis}. It will be therefore
interesting to extend our results by considering these generalized
entropies. 
\begin{figure}[H]
\caption{\protect\includegraphics[scale=0.4]{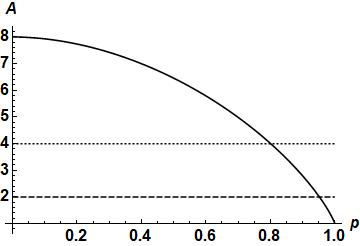}}

\raggedright{}{\footnotesize{}Plot of $A$ for the state $p\ket{GHZ}\bra{GHZ}+\frac{1-p}{8}\mathbb{I}$
as a function of $p.$ The upper dashed line represents the threshold
to detect steering, while the lower dashed line is the threshold for
GMS. }{\footnotesize \par}
\end{figure}

We acknowledge useful feedback from C. Jebaratnam and P. Skrzypczyk.

\end{document}